\documentclass[preprint,3p,12pt]{elsarticle}

\usepackage[numbers]{natbib}
\usepackage{amssymb}
\usepackage{amsmath}
\usepackage{subcaption}
\usepackage{wasysym}
\usepackage{booktabs}
\usepackage{multirow}

\journal{Speech Communication}

\begin{document}
\begin{frontmatter}

\title{AVFSNet: Audio-Visual Speech Separation for Flexible Number of Speakers with Multi-Scale and Multi-Task Learning}

\author[label1]{Daning Zhang} 
\ead{zdninger@mail.sdu.edu.cn}
\author[label1]{Yuanjie Deng} 
\ead{dengyj@mail.sdu.edu.cn}
\author[label1,label2]{Ying Wei\cormark[cor1]}
\ead{eleweiy@sdu.edu.cn}
\author[label1]{Bing Ji} 

\affiliation[label1]{
  organization={School of Control Science and Engineering, Shandong University},
  city={Jinan},
  postcode={250061}, 
  state={Shandong Province},
  country={China},
}
\affiliation[label2]{
  organization={Industrial Technology Research Institte of Shandong Province},
  city={Jinan},
  postcode={250102}, 
  state={Shandong Province},
  country={China},
}

\cortext[cor1]{Corresponding author}

\begin{abstract}
Separating target speech from mixed signals containing flexible speaker quantities presents a challenging task. While existing methods demonstrate strong separation performance and noise robustness, they predominantly assume prior knowledge of speaker counts in mixtures. The limited research addressing unknown speaker quantity scenarios exhibits significantly constrained generalization capabilities in real acoustic environments. To overcome these challenges, this paper proposes AVFSNet---an audio-visual speech separation model integrating multi-scale encoding and parallel architecture---jointly optimized for speaker counting and multi-speaker separation tasks. The model independently separates each speaker in parallel while enhancing environmental noise adaptability through visual information integration. Comprehensive experimental evaluations demonstrate that AVFSNet achieves state-of-the-art results across multiple evaluation metrics and delivers outstanding performance on diverse datasets.
\end{abstract}




\end{frontmatter}

\section{Introduction}
\label{sec:intro}
Single-channel speech separation aims to estimate individual or multiple source speech signals from a monaural mixed speech signal \cite{SeparationOverview}. It serves as a critical front-end component for numerous downstream speech processing tasks, such as automatic speech recognition, speaker verification, and speech translation \cite{SeparationAsFEnd-1,SeparationAsFEnd-2,SeparationAsFEnd-3}. In recent years, the rapid advancements in deep learning have significantly propelled progress in speech separation, leading to remarkable achievements, particularly in deterministic acoustic environments \cite{ConvTasNet, VisualVoice, Sepformer, CTCNet}.

While significant progress has been achieved in single-channel speech separation, deploying these techniques from relatively controlled environments to complex real-world scenarios remains a substantial challenge \cite{real}. Real-world acoustic scenes are frequently characterized by unknown and dynamically changing noise and reverberation \cite{denoise}. Crucially, in such scenarios, the number of speakers in mixed audio is often unknown beforehand, yet current single-channel speech separation methods primarily rely on prior knowledge of the speaker count in the mixture \cite{ConvTasNet, VisualVoice, Sepformer}. Therefore, a key challenge in this field is to develop separation methods that can both adapt to harsh acoustic conditions and flexibly handle speech mixtures with an unknown number of speakers.

To enhance model adaptability to harsh acoustic conditions and improve system robustness, multimodal fusion strategies have demonstrated clear advantages \cite{VisualVoice, AvConvTasNet}. Unimodal methods, which rely solely on audio, often experience significant performance degradation when dealing with noisy conditions or speech mixtures involving speakers whose number is unknown \cite{DeepClustering,ConvTasNet,DPTransformer}. In contrast, visual information, such as a speaker's lip movements or facial features, inherently exhibits immunity to acoustic interference. Indeed, prior research has already shown that integrating visual information into speech separation systems can significantly boost system robustness and performance under noisy conditions \cite{AvSeparation1,IIANet,CTCNet}. This potential for cross-modal information fusion enables systems to more effectively separate target speech in complex scenarios, thereby compensating for the limitations of unimodal methods in extreme conditions.

Despite the immense potential of visual information in enhancing the robustness of speech separation systems, integrating it into systems that handle an unknown number of speakers is far from trivial. From a system architecture perspective, computing the number of speakers from video streams directly by introducing modules like face detection \cite{facedetection} or lip-reading \cite{lipreading} is not advisable. Such approaches would introduce additional system complexity and error accumulation, thereby hindering overall system optimization. Therefore, it is crucial to explore feasible solutions for integrating audio-visual fusion strategies into separation tasks with an unknown number of speakers, particularly when the system cannot explicitly or implicitly acquire prior knowledge about the speaker count from its inputs.

Building upon the development of robust speech separation systems, prior research has also addressed speech separation when the number of speakers is unknown \cite{FSep-1,FSep-2,FSep-3,FSep-EDA,FSep-recursive,MultiDecoderDPRNN,EEND-SS}. However, constrained by the lack of mechanisms in current mainstream neural network architectures to generate variable outputs, most of these studies are built upon idealized assumptions, such as an upper bound on the number of speakers in the mixed audio \cite{USED, SD-SepEDA,FSep-EDA, EEND-SS}. Furthermore, employing the same network structure to learn mixed speech patterns containing varying numbers of speakers can lead to optimization difficulties \cite{EEND-SS}. Consequently, these methods struggle to genuinely adapt to dynamically changing acoustic environments. To overcome these limitations, this paper advocates for exploring novel separation strategies that more effectively align with the dynamic nature of sound sources in real-world acoustic environments.

As previously mentioned, current network architectures for unknown speaker speech separation methods exhibit limitations. Many existing approaches attempt to learn and disentangle diverse and complex mixed speech patterns using a single-scale encoding mechanism \cite{ConvTasNet, EEND-SS, Sepformer}. This inherent limitation restricts their adaptability and generalization in dynamic acoustic environments, as models with a single fixed receptive field or time window struggle to simultaneously capture both short-term and long-term acoustic features \cite{ConvTasNet,DPRNN,Sepformer,EEND-SS}. To address the complex acoustic challenges posed by dynamically changing speaker counts and real-world interference, recent speech separation research has increasingly adopted multi-scale feature modeling for long mixed speech sequences as a novel solution. Since Luo et al. introduced the Dual-Path Recurrent Neural Network (DPRNN) \cite{DPRNN}, multi-scale modeling methods, exemplified by the dual-path framework, have been widely applied in separation networks, leading to significant performance improvements \cite{AvSepformer, MFFN, DPRNN-1, SAGRNN}. Inspired by these advancements, recent work has integrated multi-scale modeling into feature encoders \cite{MultiScale-1, Ablation-A2, Conformer, Branchformer}. Most notably, Branchformer \cite{Branchformer} has demonstrated exceptional performance in speech recognition due to its flexible and adaptive design. Therefore, it is essential to integrate multi-scale modeling into unknown speaker separation. This integration promises to significantly enhance system adaptability in dynamic acoustic environments, thereby laying a solid foundation for achieving high-quality, adaptive separation.

In summary, current speech separation methods face two critical limitations: (1) most existing approaches struggle to handle dynamically varying speaker quantities, and (2) current flexible-quantity separation methods often exhibit degraded performance in noisy environments. To address these challenges, this paper introduces AVFSNet, a novel audiovisual speech separation method specifically designed for scenarios with a flexible number of speakers. AVFSNet innovatively integrates audiovisual multimodal fusion into unknown-quantity speaker separation, jointly implementing both speaker counting and multi-speaker separation tasks. Architecturally, AVFSNet achieves robust separation in dynamic unknown speaker-count scenarios through the incorporation of a Branchformer multi-scale encoder and a parallel independent separation framework. The principal contributions of this work are as follows:
\begin{enumerate}
  \item AVFSNet achieves audio-visual separation for an unknown number of speakers without relying on any visual prior information, demonstrating excellent performance in challenging mixed-speech scenarios.
  \item AVFSNet introduces a parallel separation architecture that eliminates error accumulation in conventional systems and removes the upper limit on separable speakers.
  \item AVFSNet integrates multi-speaker separation and counting within a unified multi-task framework, utilizing an innovative "separation-before-counting" process and architectural optimization for superior robustness and performance.
  \item By employing Branchformer for global-local multi-scale audio modeling, AVFSNet enhances feature representation for mixed speech with varying speaker numbers.
\end{enumerate}

\section{Related Works}
\subsection{Audio-Visual Speaker Separation}
Neuroimaging studies demonstrate that the human auditory cognition system operates as a multimodal system, where visual information significantly enhances the brain's capacity to segregate and comprehend target speech from mixed auditory inputs \cite{VisualEffect1,VisualEffect2}. Building upon this neurobiological foundation, numerous studies have incorporated visual information into speaker separation frameworks \cite{LookingSpeech,AvConvTasNet,VisualVoice,LookingtoListen,AvSeparation1}. Empirical evidence confirms that leveraging facial images \cite{VisualVoice} or lip movements \cite{MuSE, USEV, AvSepformer} to guide target speaker separation achieves superior performance. Audiovisual multimodal fusion techniques maintain robust separation capabilities even in complex acoustic environments. However, these implementations predominantly assume prior knowledge of speaker counts for model design and optimization---an assumption difficult to satisfy in real-world applications.

This work inherits the audiovisual multimodal fusion framework for target speaker separation from existing approaches. However, unlike the aforementioned methods, our model processes visual information and mixed speech signals for each potential target speaker in parallel and independently, thereby enabling individual speaker separation within mixed speech containing flexible speaker quantities.

\subsection{Speech Separation for an Unknown Number of Speakers}
Recently, the task of separating speakers from mixed speech with an unknown number of speakers has garnered significant attention from researchers \cite{FSep-1, FSep-2, FSep-3, FSep-EDA, FSep-recursive, MultiDecoderDPRNN, EEND-SS}. Evidently, in real-world acoustic environments, it is extremely common for the number of sources in mixed audio to be unknown. These studies can be broadly categorized into three approaches. 

One strategy involves recursively separating each speaker from the mixed speech \cite{FSep-recursive, FSep-2, FSep-3}. This approach operates by iteratively extracting one speaker at a time from the mixture until no more speakers are detected or a predefined stop criterion is met. For instance, Takahashi et al. \cite{FSep-recursive} propose a recursive separation framework that iteratively extracts one speaker and updates the residual signal via OR-PIT training with a binary classifier for termination. Similarly, Shi et al. \cite{FSep-3} introduce a conditional chain mapping model that generates sequences iteratively by conditioning on previous outputs, supporting variable speaker counts. While these recursive methods offer significant flexibility as they don't require pre-estimation of speaker count, they utilize residual signals from previous iterations for subsequent separation, which results in progressively degraded quality for later-separated speakers.

Another major approach focuses on estimating the number of speakers in the mixture prior to separation and then selecting the model branch corresponding to the estimated number \cite{MultiDecoderDPRNN, EEND-SS, FSep-EDA, SD-SepEDA, USED}. This strategy integrates a speaker counting module before the separation stage, allowing the separation network to be explicitly conditioned on the predicted number of active speakers. For example, \cite{MultiDecoderDPRNN} incorporates a "count-head" that treats the task of speaker counting as a multi-class classification problem. Approaches like \cite{EEND-SS, USED} combined end-to-end neural diarization (EEND) with speech separation, where EEND simultaneously estimates speaker presence and provides crucial cues for the subsequent separation network. Furthermore, approaches like \cite{FSep-EDA, SD-SepEDA} are primarily based on the attractor-decoder framework. These methods aim to accurately determine the number of active speakers by estimating an attractor for each speaker and subsequently judging the existence probability of these attractors. Unfortunately, methods within this category often employ multiple decoders to handle varying numbers of speakers and rely on a shared backbone network for learning mixed speech features. This architecture inherently limits their scalability to a wider range of speaker counts and frequently results in suboptimal overall performance.

A third distinct strategy involves separating the mixture using the maximum possible number of speakers and subsequently filtering the separated signals to identify those that actually exist \cite{FSep-1}. Naturally, such methods require setting an upper limit on the maximum number of speakers. Maximum speaker-count constrained approaches induce separation errors when actual speaker quantities exceed predefined thresholds. 

Furthermore, while the aforementioned approaches have advanced speaker separation for an unknown number of speakers, they share a common limitation: they predominantly rely on audio-only frameworks. This singular reliance on acoustic information makes them inherently susceptible to degradation in challenging low-SNR (Signal-to-Noise Ratio) scenarios, especially when confronted with various types of environmental noise. Without additional modalities or more robust noise handling mechanisms, their performance often falls short in real-world noisy conditions. 

Inspired by the aforementioned methods, our proposed approach adopts a parallel, flexible-number branch architecture. This design effectively overcomes the limitations of prior works by simultaneously preventing the accumulation of separation errors and circumventing the constraint of a predefined maximum number of speakers.

\subsection{Speaker Counting in Mixed Speech}
Speaker counting, which refers to the task of estimating the number of active speakers in continuous audio streams, constitutes a subtask of speaker diarization. Evidently, accurate speaker counting serves as a necessary front-end component for numerous speech processing tasks. The prevailing end-to-end speaker counting approaches primarily comprise regression-based counting networks, notably represented by CountNet \cite{CountNet}, and probability distribution prediction methods, exemplified by the Encoder-Decoder-Attractor (EDA) framework \cite{EEND-EDA, EEND-SS, SD-SepEDA}. Regression-based counting networks exhibit structural simplicity, computational efficiency, and the absence of upper limits on quantity estimation, yet demonstrate insufficient robustness. In contrast, the probability distribution prediction methods represented by the EDA approach have improved robustness and facilitate integration with downstream tasks \cite{EEND-SS, SD-SepEDA}, while being constrained by maximum speaker number limitations.

The proposed method also employs a probability distribution prediction methodology, generating probabilistic estimates for each potential speaker in mixed speech. However, in contrast to the aforementioned approaches, our method partially mitigates the limitation of maximum speaker numbers inherent to this category of techniques.

\section{Method}
\label{sec:method}
\subsection{Problem Formulation}
Given a video $V$ containing $M$ speakers, the single-channel linear mixture of all speakers' speech in the video is represented as $x(t)$:
\begin{equation}
  x(t) = \sum^{N}_{i=1}s_i(t)+n(t)
\end{equation}
where $s_i(t)$ denotes the source speech signal of the $i$-th speaker. $n(t)$ is the noise signal. $N$ is the number of active speakers. The lip movement of each speaker $v_i$ can be extracted from $V$, yielding the set:
\begin{equation}
  \mathcal{V}=\{v_1,v_2,\ldots,v_i,\ldots,v_M\}
\end{equation}

The objective of this work is to separate the individual speech signals $\{s_i(t)\}_{i=1}^N$ from the mixed speech signal $x(t)$, where $N$ is unknown. To simplify the problem, this paper assumes that $N$ does not exceed $M$ and that no new speakers join during the entire speaking activity.

\subsection{Overall Framwork}
\begin{figure}
  \centering
    \includegraphics[width=0.99\linewidth]{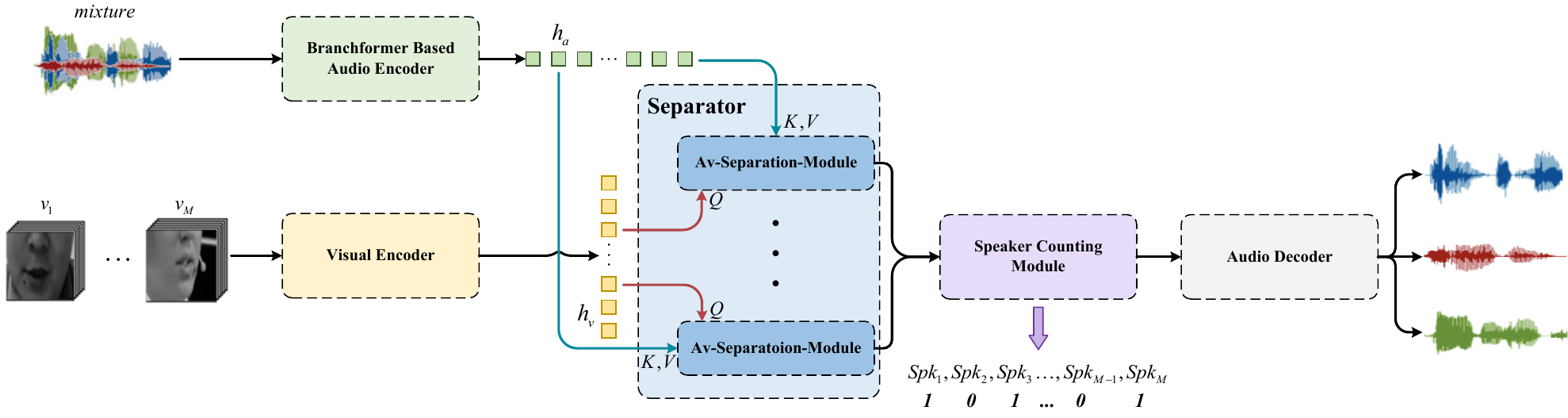}
    \caption{Overall architecture of the proposed AVFSNet. The system employs audiovisual multimodal fusion to achieve unknown-quantity speaker separation and counting through a multi-scale audio encoder (Branchformer-based) and a parallel-structured separator with independent parameter branches.}
    \label{fig:overall}
\end{figure}
The proposed model comprises an audio encoder, a visual encoder, a separator, an audio decoder, and a speaker counting module. The overall architecture is shown in Figure \ref{fig:overall}. To enable speaker-independent separation without presupposing the speaker count, we avoid providing fixed visual embeddings that might contain speaker count information. Consequently, the proposed method accepts both the lip movement sequences from all speakers in the video $V$ and the mixed speech as inputs. Instead of a conventional convolutional encoder, this work utilizes the Branchformer encoder to encode the mixed speech signal. The separator generates a mask for each potential speaker based on the audiovisual features derived by the encoders. The speaker counting module assesses the validity of the outputs of the separator and estimates the number of active speakers. Finally, the decoder reconstructs the separated signals in the time domain.

\subsection{Branchformer Based Audio Encoder}
\begin{figure}[!t]
    \centering
    \includegraphics[width=0.99\linewidth]{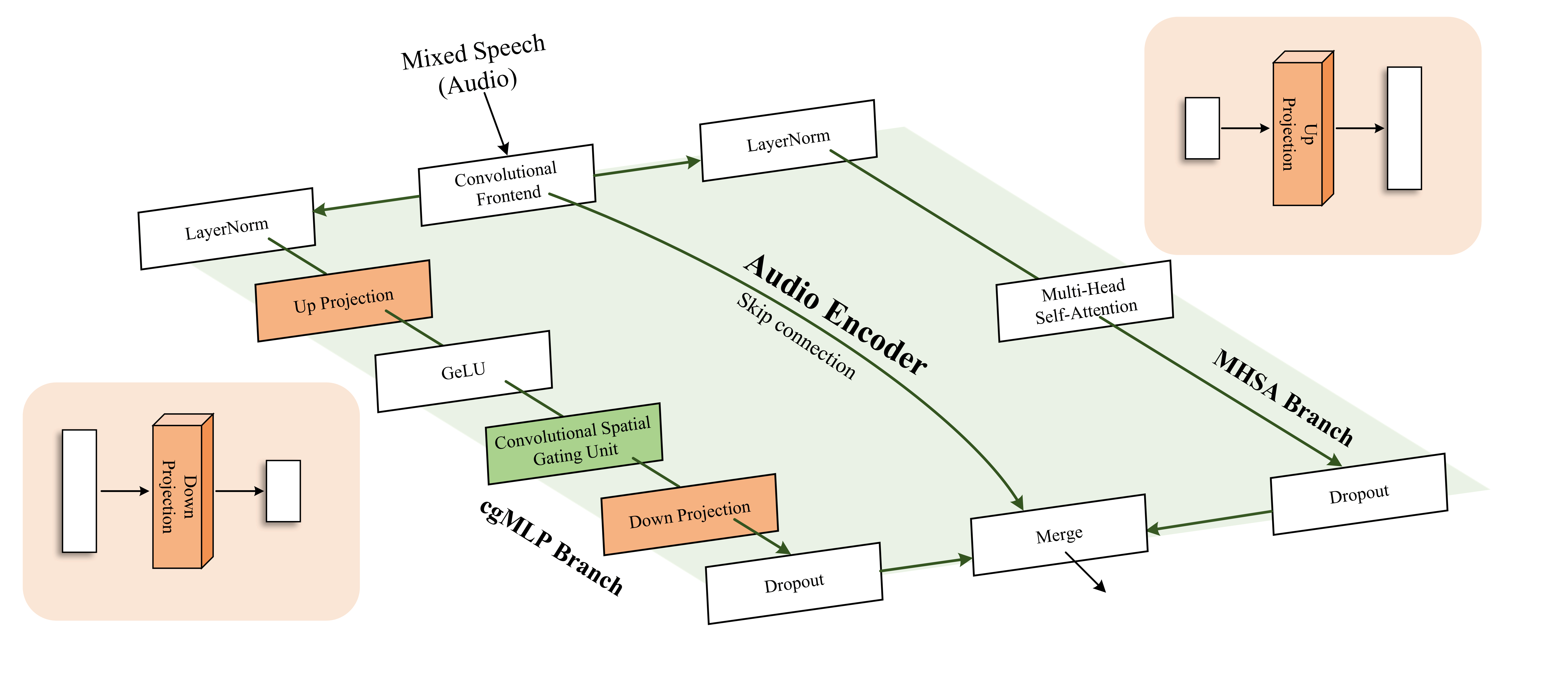}
    \caption{Proposed Branchformer-based audio encoder. The architecture contains two parallel branches: a multi-head self-attention branch and a convolutional-gated MLP branch. Outputs from both branches are fused via skip connections with the original input, yielding encoded features through weighted summation. Up projection and down projection correspond to the two learnable projection modules in the CGMLP branch. Their objective is to remap the feature dimensions of the input: the up projection expands feature dimensions, while the down projection compresses feature dimensions.}
    \label{fig:audioencoder}
\end{figure}
The conventional convolutional encoder typically employs a 1D convolutional layer with a kernel size of $K$ and a stride of $K/2$ to encode features from the mixed speech $x\in\mathbb{R}^{1\times T}$. This operation is formulated as:
\begin{equation}
  h_{ce}=\text{Conv1D}(x)\in\mathbb{R}^{D\times L}
\end{equation}
where $D$ denotes the feature dimension and $L$ represents the temporal dimension. The encoder transforms the raw speech signal into a discriminative representation that captures key information for speech separation, though without explicitly filtering noise or interference. Crucially, due to the inherently local receptive field of convolution operations, such encoders struggle to model both long-range dependencies and localized patterns simultaneously in mixed speech. Consequently, they exhibit limited representational capacity for diverse acoustic conditions and noise types.

As mentioned in Section \ref{sec:intro}, inspired by other multi-scale speech separation studies \cite{MultiScale-1, Ablation-A2}, this work proposes an encoder architecture that incorporates Branchformer \cite{Branchformer} for speech separation. This branching design leverages Branchformer to simultaneously capture global and local contextual dependencies within audio features, thereby enhancing representational capacity and robustness in complex acoustic scenes. The proposed encoder structure is illustrated in Figure \ref{fig:audioencoder}.

The convolutional frontend first processes the time-domain mixture signal $x\in\mathbb{R}^{1\times T}$ to generate an audio feature representation $h_x\in\mathbb{R}^{D\times L}$,  operating similarly to conventional convolutional encoders. This representation is then processed by two parallel encoding branches:
\begin{enumerate}
  \item A multi-head self-attention branch capturing long-range dependencies in the mixture
  \item A convolutional gating MLP (cgMLP) branch extracting local dependencies
\end{enumerate}

Features from both branches are merged via weighted averaging, and skip connections integrate the frontend's original output with the merged features. This yields the final encoded audio feature $h_a\in\mathbb{R}^{D\times L}$, where $D$ denotes the feature dimension, and $L$ represents the temporal dimension. The encoder's operational workflow is detailed as follows.

\subsubsection{Multi-Head Self-Attention Branch}
The multi-head self-attention mechanism computes correlations between each temporal position in the input sequence and all other positions. This enables comprehensive information exchange across all time steps, allowing the branch to effectively capture long-range dependencies in speech mixtures. In our multi-head self-attention branch, the input $h_x$ first undergoes normalization (LayerNorm), then is projected into query matrix $Q_x$, key matrix $K_x$, and value matrix $V_x$. The multi-head self-attention output is computed as:
\begin{equation}
  \begin{aligned}
    h_{\text{mhsa}} & =\text{MultiHeadAttention}(Q_x, K_x, V_x) \\
        & =\text{Softmax}(\frac{Q_xK_x^\top}{\sqrt{d}})V_x\in\mathbb{R}^{D\times L}
  \end{aligned}  
\end{equation}  
where $d=D/N_H$, with $N_H$ denoting the number of attention heads in the multi-head attention layer.

\subsubsection{cgMLP Branch}
The cgMLP architecture employs 1D depth-wise convolution to model local dependencies within speech mixtures, while implementing channel-wise feature modulation via a linear gating mechanism. As illustrated in Figure \ref{fig:CSGU}, its Convolutional Spatial Gating Unit (CSGU) operates through the following computational stages:

Formally, the input $h_x$ is first normalized using layer normalization and transposed. Its feature dimension is then expanded from $D$ to $D_{\text{hidden}}$ through a linear projection (Up Projection).
\begin{equation}
  h_Z=\text{GeLU}(\text{Linear}(h_x))\in\mathbb{R}^{D_{\text{hidden}}\times L}
\end{equation}

Subsequently, the CSGU partitions $h_Z$ into two feature segments $h_{Z1},h_{Z2}\in\mathbb{R}^{\frac{D_{\text{hidden}}}{2}\times L}$. The segment $h_{Z1}$ preserves original feature information through direct propagation. Concurrently, $h_{Z2}$ undergoes layer normalization followed by 1D depth-wise convolution to extract local dependencies and compute the gating signals:
\begin{equation}
  h_{Z2}'=\text{DWConv1D}(\text{LayerNorm}(h_{Z2}))\in\mathbb{R}^{\frac{D_{\text{hidden}}}{2}\times L}
\end{equation}

The linear gating mechanism is implemented through element-wise multiplication of the two segments:
\begin{equation}
  h_{Z}'=h_{Z1} \otimes h_{Z2}'\in\mathbb{R}^{\frac{D_{\text{hidden}}}{2}\times L}
\end{equation}

Finally, the original dimensions are restored through linear projection (Down Projection) and transposition:
\begin{equation}
  h_{\text{cgMLP}}=\text{Linear}(h_{Z}')\in\mathbb{R}^{D\times L}
\end{equation}

\subsubsection{Merging Branches}
The proposed architecture utilizes a weighted summation scheme to fuse branch features:
\begin{equation}
  h_a=\alpha h_{\text{mhsa}}+(1-\alpha)h_{\text{cgMLP}}\in\mathbb{R}^{D\times L}
\end{equation}
where $\alpha$ is a learnable hyperparameter governing the fusion ratio between global and local features. This adaptive merging dynamically balances the contributions of global and local dependencies to the final encoded output while reducing cross-branch interference.

\subsection{Visual Encoder}
\begin{figure}[!t]
    \centering
    \begin{minipage}[c]{0.46\linewidth}
      \includegraphics[width=0.99\linewidth]{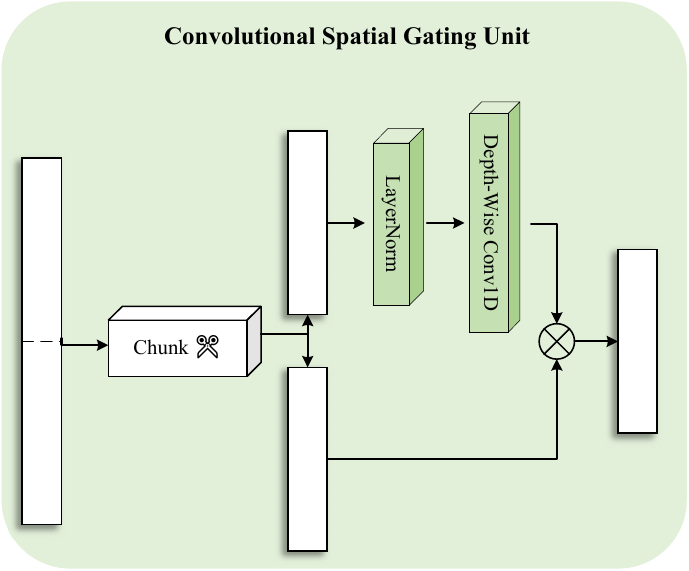}
      \caption{Convolutional spatial gating unit in audio encoder. The component computes gating signals from projected audio features and executes linear gating operations.}
      \label{fig:CSGU}
    \end{minipage}
    \hfill
    \begin{minipage}[c]{0.50\linewidth}
      \includegraphics[width=0.99\linewidth]{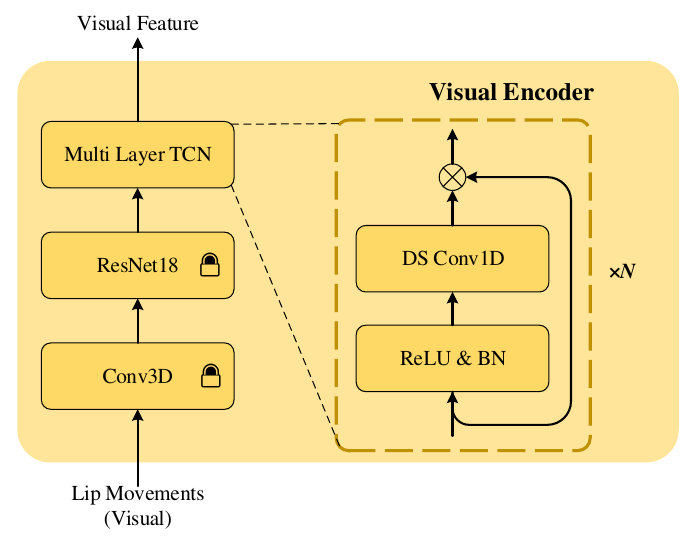}
      \caption{Schematic of visual encoder}
      \label{fig:visualencoder}
    \end{minipage}
\end{figure}
The visual encoder follows established designs for audio-visual target speaker extraction \cite{AvConvTasNet, MuSE, USEV, AvSepformer}. As depicted in Figure \ref{fig:visualencoder}, it comprises two components: a lip embedding extractor and a multilayer Temporal Convolutional Network (TCN) module. The lip embedding extractor comprises a Conv3D layer and a ResNet18 network, which has been fully pre-trained on lip-reading tasks. The lip movements $v_i$ of the speaker $i$ are first processed by the lip embedding extractor and then passed through the multilayer TCN to obtain visual features $h_{v,i}\in\mathbb{R}^{D\times I}$, where $D$ denotes the feature dimension and $I$ represents the temporal dimension of the visual features.

\subsection{Separator}
\begin{figure}
  \centering
  \includegraphics[width=0.99\linewidth]{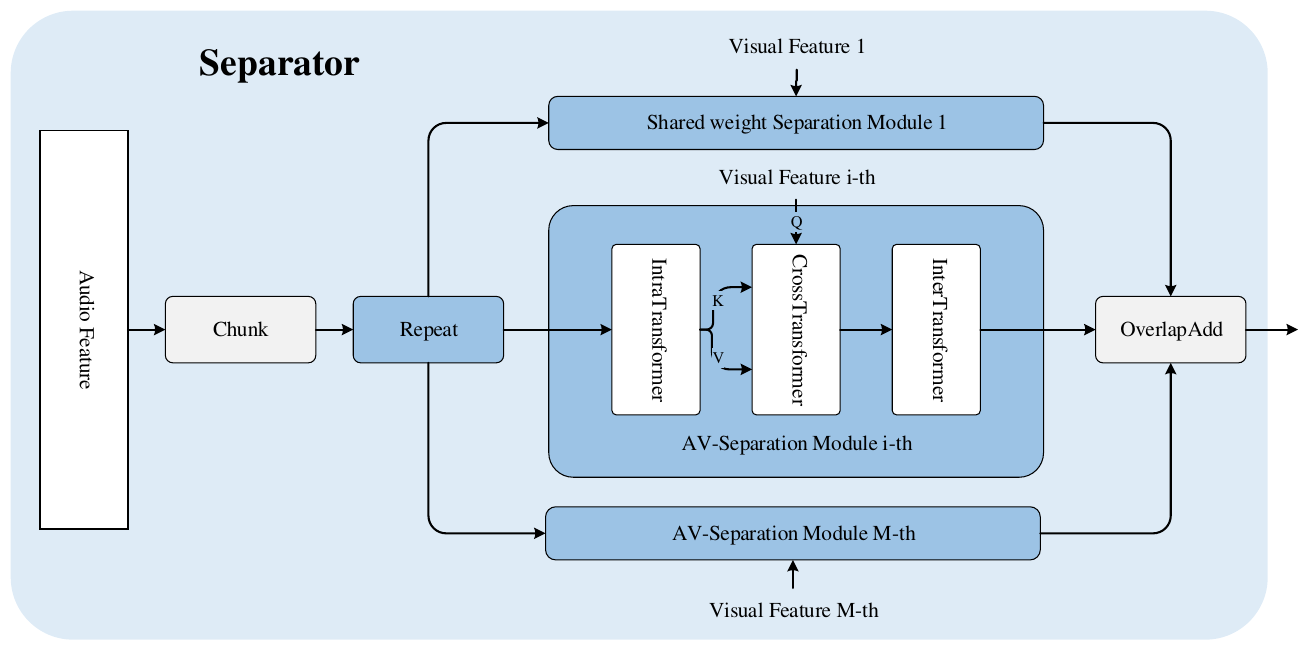}
  \caption{Separator architecture schematic. The parallel-structured separator employs flexible-quantity branches where each audiovisual separation module extracts one potential target speaker from mixed speech via visual cues. Each branch integrates multiple IntraTransformer and InterTransformer blocks, achieving audiovisual fusion through cross-attention mechanisms between spectral embeddings and lip movement features.}
  \label{fig:separator}
\end{figure}
Figure \ref{fig:separator} illustrates the architecture of our separator. The separator utilizes multiple parallel audio-visual separation modules with shared weights to separate each potential speaker. Given the mixed speech features encoded by the audio encoder, the separator first segments these features into overlapping chunks, replicates the chunked sequences, and distributes them across all separation modules. Each module then independently estimates a chunk-level mask for its target speaker, guided by the corresponding visual features, and reconstructs the full-length mask via overlap-add operations.

\subsubsection{Chunk, repeat, and overlap-add}
During the chunking process, the input 2D audio feature 
$h_a$ is padded and partitioned into overlapping segments of length $C$ with a hop size of $C/2$. These segments are concatenated to form a 3D chunk-wise tensor $h_a'\in\mathbb{R}^{D\times C\times I}$, where $I$ denotes the number of chunks. By configuring chunking parameters to match the temporal resolution of visual features, the chunk count is precisely aligned with the temporal dimension I of $h_{v,i}$.

According to the number of potential target speakers $M$, the chunked audio features $h_a'$ are replicated $M$ times and distributed to $M$ parallel branches, while shared-weight audio-visual separation modules are instantiated accordingly. This architecture leverages parallel branches operating independently to process each potential speaker's signal:
\begin{equation}
  \text{for}\enspace i=1\enspace \text{to}\enspace\text{M}:\hat{m}_i'=\text{SeparationModule}_i(h_a',h_{v,i})\in\mathbb{R}^{D\times C\times I}
\end{equation}

After estimating the chunked 3D mask in the audio-visual separation module, an overlap-add operation --- the functional inverse of chunking --- reconstructs the continuous 2D target mask $\hat{m}_i\in\mathbb{R}^{D\times L}$. 

\subsubsection{Separation Module}
This work adopts the widely used AV-Sepformer \cite{AvSepformer} architecture as the backbone of our audio-visual separation module. Sepformer \cite{Sepformer} is a time-domain single-channel speech separation approach that leverages a dual-path Transformer to model intra-chunk and inter-chunk dependencies within segmented audio features, capturing both short- and long-term acoustic patterns. AV-Sepformer extends this framework by integrating a cross-attention mechanism for audio-visual feature fusion. This design ensures temporal alignment between modalities while unifying fusion and target speech extraction within the attention paradigm.

The proposed separation module primarily comprises three components: IntraTransformer, CrossModalTransformer, and InterTransformer. The IntraTransformer performs intra-chunk modeling via self-attention along the dimension $I$:
\begin{equation}
  h_a''[:,:,i]=\text{IntraTransformer}(h_a'[:,:,i]), i\in\{1,\ldots,I\}
\end{equation}

Then, the CrossModalTransformer Fuses modalities by computing cross-attention with visual features as queries (Q) and audio features as keys (K)/values (V):
\begin{equation}
  h_f=\text{Softmax}(\dfrac{h_{v,i}W_v^Q\cdot(h_a''W_a^K)^\top}{\sqrt{D}})h_a''W_a^V\in\mathbb{R}^{D\times C\times I}
\end{equation}

Finally, the InterTransformer models inter-chunk dependencies along the dimension $C$ to generate chunk-wise masks $\hat{m}'\in\mathbb{R}^{D\times C\times I}$:
\begin{equation}
  \hat{m}'[:,i,:]=\text{InterTransformer}(h_f[:,i,:]), i\in\{1,\ldots,C\}
\end{equation}

\subsection{Audio Decoder}
The decoder reconstructs the time-domain waveform of the separated speech by applying a 1D transposed convolutional layer to the element-wise product between the mask ${\hat{m}}_i$ (generated by the separation network) and the encoded audio feature $h_a$ (output by the encoder). This transformation is formulated as:
\begin{equation}
  \hat{s}_i=\text{conv1d-transpose}(\hat{m}_i\odot h_a)
\end{equation}
where ${\hat{s}}_i$ denotes the estimated waveform of the $i$-th separated source.

\subsection{Speaker Counting Module}
\begin{figure}
    \centering
    \includegraphics[width=0.80\linewidth]{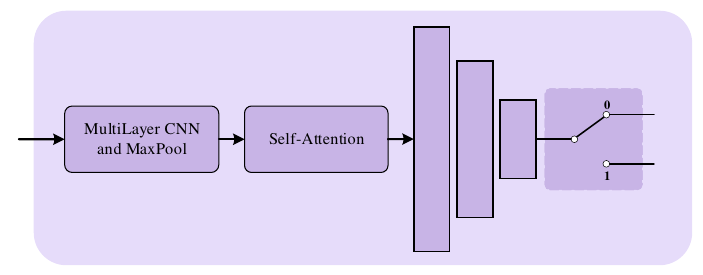}
    \caption{Schematic of speaker counting module}
    \label{fig:speakercounting}
\end{figure}

In the proposed architecture, each branch of the separator independently estimates masks for potential speakers. These masks comprise two categories: valid masks corresponding to target speakers present in the mixture, and phantom masks representing non-existent speakers. Evidently, when sufficiently trained, the separator naturally produces statistically distinguishable characteristics between these categories. Consequently, speaker counting reduces to a binary classification problem on the separator's mask outputs, where the counting module discriminates between valid and phantom masks. The architecture of the speaker counting module is illustrated in Figure \ref{fig:speakercounting}. It accepts separator outputs and estimates speaker existence probability $p\in[0,1]$ for each branch-wise mask.

For each input mask $\hat{m}_i\in\mathbb{R}^{D\times L}$ from the separator, the module learns features that distinguish between valid and phantom masks and maps them to the corresponding speaker existence probability $p\in[0,1]$. Initially, the input masks are transformed into frame-level feature representations using convolution and max-pooling operations. Subsequently, multi-head self-attention is applied to the resulting feature sequence to capture global dependencies and contextual information. The attended features are then compressed into a fixed-dimensional embedding vector. Finally, this vector is processed by a multilayer perceptron (MLP) to yield the estimated probability:
\begin{equation}
  p=\text{Sigmoid}(\text{Linear}_2(\text{Linear}_1(f)))\in[0, 1]
\end{equation}

\section{Experiments}
\subsection{Datasets}
\subsubsection{VoxCeleb2}
VoxCeleb2 \cite{VoxCeleb2} comprises approximately 1 million speech clips, totaling over 2,000 hours. It includes nearly 6,000 speakers with diverse ages, genders, accents, and nationalities, and captures complex noise scenarios.
In this study, 800 speakers from the VoxCeleb2 training partition were selected for model training. From the test partition, 118 speakers were further divided into validation and test subsets, each containing 59 speakers. Crucially, there is no overlap between speakers in the training and evaluation sets. All selected utterances exceeded 4 seconds in duration. We generated 20,000, 5,000, and 3,000 mixtures for the training, validation, and test sets, respectively. These mixtures consisted of 2-speaker (VoxCeleb2-2mix) and 3-speaker (VoxCeleb2-3mix) combinations. To assess model performance under conditions with an unknown number of speakers, we created an additional VoxCeleb2-2\&3mix test set containing both 2-speaker and 3-speaker mixtures. Consistent with prior configurations, interfering speech was mixed with the target speech at random SNRs ranging from -10 dB to 10 dB. Synchronized audio-visual data was maintained at an audio sampling rate of 16 kHz and a video frame rate of 25 FPS.

\subsubsection{LRS2}
The LRS2 dataset \cite{LRS2} comprises an audio-visual speech recognition corpus sourced from BBC television programs, containing over 140,000 video clips representing multiple languages and accents. Each sample is annotated with precise speech transcriptions and temporal alignment information, rendering it suitable for diverse research domains such as speech recognition, speech separation, and multimodal learning.

Each video clip in LRS2 features synchronized audio-visual signals from a single speaker. A subset of videos was first selected, and their video frames were separated from the corresponding audio tracks. Following the methodology employed in partitioning our VoxCeleb2 test sets, we subsequently created two test datasets: LRS2-2mix and LRS2-3mix, each containing 3,000 mixed speech samples for the 2-speaker and 3-speaker scenarios, respectively.

\subsection{Training and inference procedures}
\subsubsection{Training stages}
\label{subsec:trainingstages}
The training of the proposed model follows a three-stage strategy. We first conduct independent pre-training of the backbone network and the speaker counting module, which is motivated by the speaker counting module's dependence on the separator's separation quality. Subsequently, we conduct end-to-end joint optimization of the entire system.

During the first stage, the backbone network---comprising the audio encoder, visual encoder, separator, and audio decoder---is trained. Critically, the objective of this stage is to enable the backbone network to accurately extract target speech from mixtures of varying speaker counts by leveraging visual cues. Consequently, since only one target speaker is specified during this stage, a single separation branch is implemented in the separator. 

To ensure the model's robust generalization to diverse and more complex acoustic scenarios, it is essential to incorporate mixed speech from different numbers of speakers into the training process. For this purpose, two main strategies were explored: directly combining datasets with varying speaker counts and a transfer learning approach. Our experiment revealed that directly mixing datasets resulted in suboptimal performance (Table \ref{tab:transfer}). Consequently, a transfer learning strategy was adopted, where the model is initially trained on a two-speaker mixture dataset and subsequently fine-tuned on datasets containing more speakers. This sequential training paradigm enables the model to efficiently adapt to the complexities of higher speaker counts while retaining the fundamental separation knowledge acquired from simpler scenarios.
\begin{table}
  \centering
    \begin{tabular}{lcccccc}
        \toprule
        \multirow{2}[0]{*}{} & \multicolumn{3}{c}{LRS2} & \multicolumn{3}{c}{VoxCeleb2} \\
          & SI-SDRi & SDRi  & PESQ  & SI-SDRi & SDRi  & PESQ \\
        \midrule
        Combining & 6.43  & 6.55  & 1.29  & 6.36  & 6.64  & 1.23 \\
        Transfer  & \textbf{13.56} & \textbf{13.96} & \textbf{1.83} & \textbf{12.95} & \textbf{13.49} & \textbf{1.89} \\
        \bottomrule
    \end{tabular}
    \caption{Comparison of two training methods. Combining means randomly mixing 2-speaker and 3-speaker mixed speech samples. Transfer means the model is first trained on 2-speaker mixed speech samples and then fine-tuned on 3-speaker mixed speech samples.}
    \label{tab:transfer}
\end{table}

During the second stage, the backbone network trained and frozen from the first stage remains fixed while the speaker counting module undergoes isolated training. This module is trained using both positive and negative audiovisual sample pairs. Each sample pair consists of:
\begin{enumerate}
  \item A mixed speech signal, and
  \item A speaker lip sequence image serving as the target speaker extraction cue
\end{enumerate}

Positive pairs indicate the presence of the target speaker (corresponding to the visual cue) within the mixed speech, whereas negative pairs indicate their absence. Critically, the separation outputs generated by the frozen backbone network exhibit a significant divergence between these two sample types. The training objective of the speaker counting module is to capture this divergence, thereby enabling effective detection of valid target speech presence and accurate speaker counting.

During the third stage, the entire network, initialized with pre-trained weights from preceding stages, undergoes fine-tuning using a joint loss function.

\subsubsection{Loss Functions}
Different loss functions are employed at each training stage to align with their distinct objectives. During the first stage, we optimize the backbone network using the scale-invariant signal-to-noise ratio (SI-SNR) \cite{SISNR}, defined as:
\begin{equation}
  \mathcal{L}_{\text{SISNR}}=20\log_{10}(\dfrac{\vert\vert\alpha\cdot s\vert\vert}{\vert\vert\hat{s}-\alpha\cdot s\vert\vert})
\end{equation}
where $\alpha$ denotes the optimal scaling factor for eliminating amplitude discrepancies between signals, defined as:
\begin{equation}
  \alpha=\dfrac{\hat{s}^\top s}{s^\top s}
\end{equation}
where $\hat{s}$ and $s$ denote the estimated and original clean speech signals, respectively.

During the second stage, we compute the cross-entropy loss between the ground-truth speaker presence labels $y$ and predicted probabilities $p$ for each audio-visual sample:
\begin{equation}
  \mathcal{L}_{\text{CrossEntropy}}=-y\log(p)-(1-y)\log(1-p)
\end{equation}

In the final stage, we adopt the joint loss function \cite{JointLoss}:
\begin{equation}
  \mathcal{L} = \dfrac{1}{2\sigma_1^2}\mathcal{L}_{\text{SISNR}}+\dfrac{1}{2\sigma_2^2}\mathcal{L}_{\text{CrossEntropy}}+\log\sigma_1+\log\sigma_2
\end{equation}
where $\sigma_1$ and $\sigma_2$ are learnable parameters dynamically weighting the contribution of each loss component.

\subsubsection{Inference stage}
During inference, the model receives the lip movements of all detected speakers within a video segment along with the mixed audio input. This process yields a set of estimated target speech signals $S_{out}=\{s_1,\ldots,s_M\}$ nd their corresponding presence probabilities $P=\{p_1,\ldots,p_M\}$. The final output sources $\hat{S}$ are then selected from $S_{\text{out}}$ by retaining only those speech signals whose presence probability meets or exceeds a predefined threshold $\tau$:
\begin{equation}
  \hat{S}=\max\{s_i\in S_{\text{out}} | p_i \geq\tau\}
\end{equation}

\subsection{Implementation Details}
\begin{table}
    \centering
    \begin{tabular}{lcc}
        \toprule
        Symbol & Description & Value \\
        \midrule
        $K$ & Kernel size in Conv1D front-end of audio encoder & 16 \\
        $C$ & Length of the chunk & 160 \\
        $N_{\text{intra}}$ & Number of the IntraTransformer & 8 \\
        $N_{\text{inter}}$ & Number of the InterTransformer & 7 \\
        $D$ & Feature dimensionality & 256 \\
        $D_{\text{hidden}}$ & Feature dimensionality after projection & 2048 \\
        $N_H$ & Number of head in multi-head attention & 8 \\
        \bottomrule
    \end{tabular}
    \caption{Hyperparameters of AVFSNet.}
    \label{tab:hyperparam}
\end{table}

We implemented the AVFSNet system as detailed in Section \ref{sec:method}, with corresponding hyperparameters presented in Table \ref{tab:hyperparam}. $C$ was set to 160 to align the dimensionality $I$ of both chunked audio and visual features. Adam was employed as the optimizer across all training stages.

The training stages were conducted on two NVIDIA GeForce RTX 3090 GPUs. To accelerate the training process, we leveraged Distributed Data Parallel (DDP) and Automatic Mixed Precision (AMP). Throughout the entire training regimen, we ensured convergence stability through adaptive learning rate scheduling based on validation set performance and gradient clipping. Concurrently, we implemented an early stopping strategy to prevent overfitting.

\subsection{Main Comparison Methods}
AV-DPRNN \cite{DPRNN} is an audio-visual model for speech separation and target speaker extraction, utilizing a dual-path RNN architecture. By integrating visual cues into the dual-path processing framework, it achieves consistent separation performance in multi-speaker scenarios and attains state-of-the-art(SOTA) results on established benchmarks, establishing its status as a widely adopted foundational model in the field. We reimplemented this architecture and fine-tuned it on the 3-speaker dataset using transfer learning configurations identical to those applied for AVFSNet.

MuSE \cite{MuSE} employs a temporal convolutional network, enhancing the Av-ConvTasNet architecture through a self-enrollment strategy for speaker embedding learning during training. Unlike conventional speaker separation/extraction models, it introduces an auxiliary classification loss to refine the speaker representation comprehension. This auxiliary supervision enables the system to achieve SOTA performance with enhanced cross-dataset robustness.

AV-Sepformer \cite{AvSepformer} leverages a dual-path Transformer framework for multimodal separation/extraction, utilizing cross-modal cross-attention mechanisms for audio-visual fusion. The system ensures precise temporal granularity alignment between modalities and attains SOTA performance. For the audio-visual separation module in our proposed AVFSNet, we adopted AV-Sepformer as the backbone network. We reimplemented this model and fine-tuned it on the 3-speaker dataset using transfer learning configurations identical to those applied for AVFSNet.

SEANet \cite{SEANet} proposes a novel audio-visual speech separation and extraction framework. It introduces a selective auditory attention mechanism that suppresses interference via noise-oriented reverse attention, enabling robust target extraction in noisy environments. By directly estimating and subtracting interference signals, SEANet achieves SOTA performance across diverse datasets.

\subsection{Metrics}
Following previous studies, this work employs scale-invariant signal-to-distortion ratio (SI-SDR) \cite{SISNR}, signal-to-distortion ratio (SDR), and their relative improvement metrics (SI-SDRi and SDRi) as the primary metrics for evaluating the quality of separated speech. These relative metrics quantify the enhancement in the separated speech signal compared to the original mixture input. The units for all metrics are decibels (dB). Additionally, this work introduces PESQ \cite{PESQ} (Perceptual Evaluation of Speech Quality) and STOI \cite{STOI} (Short-Term Objective Intelligibility) to further assess speech quality. Higher values for all these metrics indicate better performance.

It should be noted that during the ablation experiment phase, the model may misjudge a speaker's active status. Specifically, this error occurs when the model either estimates a silent signal for an active speaker or falsely identifies a silent speaker in mixed speech as active. These misjudgments cause the reference or estimated signals for the corresponding speaker in the evaluation metric computation to become all-zero. This can result in ill-defined or unreliable metric values. To impose an error penalty and ensure robust evaluation, we set the corresponding speaker's evaluation metric to 0 when such an incorrect judgment occurs, given that all metrics in this paper are inherently positive.

\section{Results and Analysis}
This section presents the results, organized into four subsections: Subsection \ref{subsec:evaluation} compares AVFSNet with previous SOTA methods on the VoxCeleb2 and LRS2 datasets. Subsection \ref{subsec:ablation} conducts the ablation study to validate the effectiveness of AVFSNet's architecture. Subsection \ref{subsec:quality} empolys the quality study to prove the robustness of AVFSNet with different evaluation conditions and settings. Finally, Subsection \ref{subsec:visualization} provides some visualization results.

\subsection{Evaluation}
\label{subsec:evaluation}
\subsubsection{Evaluation on 2-speaker mixed speech dataset}
Table \ref{tab:EVSpk2} compares the proposed AVFSNet with previous methods on two-speaker mixed speech test sets from VoxCeleb2 and LRS2. On the LRS2-2mix dataset, AVFSNet achieves the best performance across all metrics (SI-SDRi, SDRi, and PESQ). Specifically, it outperforms the baseline AV-Sepformer by 1.7 dB in SI-SDRi (14.34 dB vs. 12.64 dB) and 1.16 dB in SDRi (14.13 dB vs. 12.97 dB). Compared to SEANet (selective auditory attention) and MuSE (speaker embeddings), AVFSNet demonstrates significant improvements: SI-SDRi values of 14.34 dB vs. 13.05 dB (SEANet) and 11.01 dB (MuSE), and PESQ scores of 2.51 vs. 2.25 (SEANet) and 1.98 (MuSE). This superiority in target speech separation is attributed to AVFSNet's multiscale audio feature extraction and cross-attention-based audio-visual fusion.

On the more challenging VoxCeleb2-2mix dataset with diverse background noise and complex acoustic scenes, all models exhibit performance degradation relative to LRS2-2mix. Nevertheless, AVFSNet maintains a competitive advantage. While AVFSNet surpasses SEANet in SI-SDRi (12.93 dB vs. 12.77 dB), it shows a slight deficit in SDRi (12.97 dB vs. SEANet's 13.13 dB). This discrepancy likely stems from SEANet's selective auditory attention mechanism, which enhances acoustic interference suppression. AVFSNet achieves comparable noise robustness through joint modeling of global and local contextual information.

\begin{table}
    \centering
    \begin{tabular}{c|ccc|ccc}
        \toprule
        \multirow{2}[0]{*}{Method} & \multicolumn{3}{c|}{LRS2} & \multicolumn{3}{c}{VoxCeleb2} \\
         & SI-SDRi & SDRi  & PESQ  & SI-SDRi & SDRi  & PESQ \\
        \midrule
        AV-ConvTasNet\cite{AvConvTasNet} & 10.81 & -- & --  & 10.641 & -- & 1.973 \\
        AV-DPRNN\cite{DPRNN} & 10.29 & 10.65 & 1.88  & 11.02 & 11.35 & 2.09 \\
        MuSE\cite{MuSE}  & 11.01 & 11.41 & 1.98  & 10.78 & 10.51 & 2.201 \\
        AV-Sepformer\cite{AvSepformer} & 12.64 & 12.97 & 2.2   & 11.96 & 12.29 & 2.25 \\
        SEANet\cite{SEANet} & 13.05 & 13.52 & 2.25  & 12.77 & \textbf{13.13} & 2.34 \\
        \textbf{AVFSNet} & \textbf{14.34} & \textbf{14.13} & \textbf{2.51} & \textbf{12.93} & 12.97 & \textbf{2.49} \\
        \bottomrule
    \end{tabular}
    \caption{Comparison of AVFSNet and previous methods on the LRS2-2mix and VoxCeleb2-2mix. }
    \label{tab:EVSpk2}
\end{table}

\subsubsection{Evaluation on 3-speaker mixed speech dataset}
Table \ref{tab:EVSpk3-1} presents a comparison of the proposed AVFSNet with existing methods on three-speaker mixed speech test sets derived from VoxCeleb2 and LRS2. When evaluating models trained solely on two-speaker mixtures against these three-speaker test data, all methods, as expected, exhibit noticeable performance degradation due to this training-test mismatch. Despite this domain shift, AVFSNet achieves the best performance on both the LRS2-3mix and VoxCeleb2-3mix datasets. This is attributed to AVFSNet's architecture, which employs a multi-scale modeling approach combining global and local contextual information from mixed speech, thereby enabling its high adaptability in complex acoustic environments.

To further accurately evaluate model performance on three-speaker separation tasks, as described in Subsection \ref{subsec:trainingstages} and Table \ref{tab:transfer}, we adopt a transfer learning strategy. This involves first pre-training the model on two-speaker mixed speech datasets and then fine-tuning it on three-speaker mixed speech datasets. For a comprehensive comparison, AV-DPRNN and AV-Sepformer were implemented with identical transfer fine-tuning configurations. The performance evaluation of models before and after transfer learning is presented in Table \ref{tab:EVSpk3-2}. All fine-tuned models exhibit noticeable performance improvements over their baseline counterparts trained solely on two-speaker mixtures (e.g., AVFSNet VoxCeleb2 SI-SDRi: 12.95 dB vs. 6.59 dB; LRS2 SI-SDRi: 13.56 dB vs. 6.81 dB), further confirming the necessity and effectiveness of the fine-tuning strategy.

As shown in Table \ref{tab:EVSpk3-2}, under identical conditions, the proposed method significantly outperforms the comparison models. For example, on the VoxCeleb2 dataset, our method achieves a 12.95 dB SI-SDRi—surpassing the backbone AV-Sepformer (11.59 dB) by 1.36 dB—and a 13.49 dB SDRi, exceeding AV-Sepformer's 12.15 dB by 1.34 dB. Furthermore, AVFSNet attains greater performance gains via transfer fine-tuning than AV-DPRNN and AV-Sepformer. This further confirms our architecture's superior adaptability in this optimization task, attributed to its stronger feature capture capabilities and more flexible architecture.

\begin{table}
    \centering
    \begin{tabular}{c|ccc|ccc}
        \toprule
        \multirow{2}[0]{*}{Method} & \multicolumn{3}{c|}{LRS2} & \multicolumn{3}{c}{VoxCeleb2} \\
          & SI-SDRi & SDRi  & PESQ  & SI-SDRi & SDRi  & PESQ \\
        \midrule
        AV-DPRNN\cite{DPRNN} & 5.80 & 5.98  & 1.29  & 5.74  & 5.95  & 1.35 \\
        MuSE\cite{MuSE}  & 5.63 & 5.83  & 1.27  & 5.37  & 5.62  & 1.33 \\
        AV-Sepformer\cite{AvSepformer} & 6.49  & 6.65  & 1.35  & 6.43  & 6.70   & 1.42 \\
        SEANet\cite{SEANet} & 6.77  & 6.98  & 1.35 & 6.52  & 6.85  & 1.42 \\
        \textbf{AVFSNet} & \textbf{6.81} & \textbf{7.04} & \textbf{1.36} & \textbf{6.59} & \textbf{6.9} & \textbf{1.44} \\
        \bottomrule
    \end{tabular}
    \caption{Comparison of AVFSNet and previous methods on the LRS2-3mix and VoxCeleb2-3mix.}
    \label{tab:EVSpk3-1}
\end{table}
\begin{table}
  \centering  
  \begin{tabular}{l|l|ccc|ccc}
    \toprule
    \multirow{2}{*}{Datasets}  & \multirow{2}{*}{Methods} & \multicolumn{3}{c|}{Non-Transfer} & \multicolumn{3}{c}{Transfer} \\
                               & & SISDRi & SDRi & PESQ & SISDRi & SDRi & PESQ \\
    \midrule
    \multirow{3}{*}{LRS2}      & AV-DPRNN\cite{DPRNN} & 5.80 & 5.98 & 1.29 & 9.75 & 10.28 & 1.46 \\
                               & AV-Sepformer\cite{AvSepformer} & 6.49 & 6.65 & 1.35 & 12.22 & 12.63 & 1.70 \\
                               & AVFSNet & 6.81 & 7.04 & 1.36 & \textbf{13.56} & \textbf{13.96} & \textbf{1.83} \\
    \midrule
    \multirow{3}{*}{VoxCeleb2} & AV-DPRNN\cite{DPRNN} & 5.74 & 5.95 & 1.35 & 9.70 & 10.28 & 1.53 \\
                               & AV-Sepformer\cite{AvSepformer} & 6.43 & 6.70 & 1.42 & 11.59 & 12.15 & 1.76 \\
                               & AVFSNet & 6.59 & 6.90 & 1.44 & \textbf{12.95} & \textbf{13.49} & \textbf{1.89} \\
    \bottomrule
  \end{tabular}
  \caption{Comparison of AVFSNet and previous methods on LRS2-3mix and VoxCeleb2-3mix: performance before and after transfer learning.}
  \label{tab:EVSpk3-2}
\end{table}

\subsection{Ablation Study}
\label{subsec:ablation}
\begin{figure}
    \centering
    \includegraphics[width=0.99\linewidth]{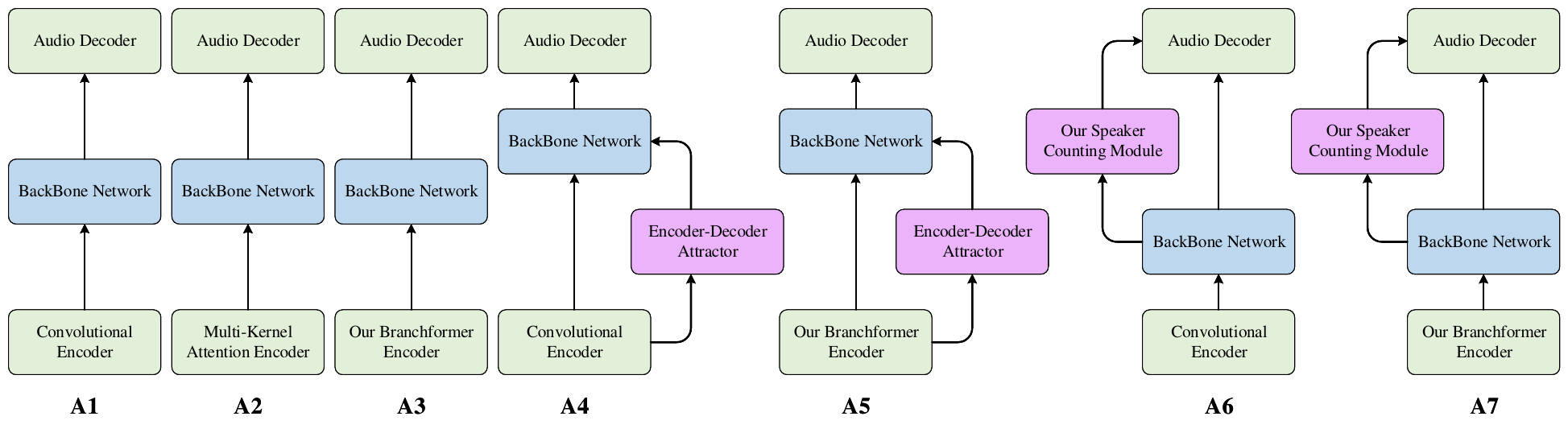}
    \caption{Seven kinds of architecture for ablation study. A1 incorporates the traditional convolutional audio encoder; A2 and A3 implement two multi-scale audio encoders building upon the backbone network; A4--A7 comprise four architectural variants combining two types of audio encoders with two speaker counting modules.}
    \label{fig:ablationstudy}
\end{figure}
In this section, ablation studies are conducted to validate the superiority of the proposed architecture and investigate its mechanisms. As shown in Figure \ref{fig:ablationstudy}, seven architectural variants (A1--A7) are constructed for ablation analysis. A1--A3 respectively utilize: (1) a conventional convolutional encoder, (2) a multi-scale encoder with multi-kernel attention \cite{Ablation-A2}, and (3) the proposed Branchformer-based multi-scale encoder. For A4--A7, two encoder types (conventional convolutional vs. proposed Branchformer-based) are combined with two speaker counting methodologies (encoder-decoder attractor-based counting module vs. proposed counting module). A4 can be considered the baseline architecture, while A7 is the proposed AVFSNet. All aforementioned architectures are implemented under identical experimental environments and configurations, with results summarized in Table \ref{tab:ablation1} and Table \ref{tab:ablation2}.

As shown in Table \ref{tab:ablation1}, on the VoxCeleb2-2mix and VoxCeleb2-3mix datasets, multi-scale encoders (A2 and A3) outperform the conventional convolutional encoder (A1) in separation performance. This confirms that multi-scale modeling enhances feature representation capacity, leading to overall performance improvements. The Branchformer-based encoder surpasses the multi-kernel attention variant (A2), indicating that while conventional multi-scale encoders capture features at varying granularities, they insufficiently exploit global contextual dependencies in speech signals. The proposed Branchformer-based architecture significantly enhances performance by synergistically integrating global and local information.

\begin{table}
    \centering
    \begin{tabular}{ccccccc}
        \toprule
        \multirow{2}[0]{*}{Method} & \multirow{2}[0]{*}{FlexNumSpk} & \multicolumn{2}{c}{VoxCeleb2-2mix} & \multicolumn{2}{c}{VoxCeleb2-3mix} \\
          & & SI-SDRi & STOI  & SI-SDRi & STOI \\
        \midrule
        A1 & \Circle & 11.96 & 0.88 & 10.68 & 0.76 \\
        A2 & \Circle & 12.03 & 0.85 & 10.95 & 0.73 \\
        A3 & \Circle & 12.93 & 0.89 & 12.95 & 0.81 \\
        \bottomrule
    \end{tabular}
    \caption{Ablation study for A1--A3 structures. \Circle denotes that the architecture \textbf{cannot} handle a flexible number of speakers.}
    \label{tab:ablation1}
\end{table}

Furthermore, as evidenced in Table \ref{tab:ablation2}, with identical speaker counting strategies (A4/A5 vs. A6/A7), architectures employing the Branchformer-based encoder consistently outperform conventional convolutional counterparts on the VoxCeleb2-2\&3mix datasets. This demonstrates that our multi-scale encoder and dynamic parallel architecture enable enhanced adaptability to complex acoustic environments.

As shown in Table \ref{tab:ablation2}, under identical encoder configurations (A4/A6 vs. A5/A7), the proposed speaker counting method achieves noticeably higher Speaker Count Accuracy (SCA) versus the encoder-decoder attractor-based approach on VoxCeleb2-2\&3mix. This disparity arises because the attractor-based method directly estimates speaker attractors and their presence probabilities from mixed speech, where cross-speaker interactions degrade performance in noisy conditions or multi-speaker scenarios. Conversely, our novel module independently estimates each potential speaker's presence probability from speaker masks, effectively eliminating inter-speaker interference and reducing task complexity.

Finally, beyond the independent contributions of each module, the performance of the complete AVFSNet (A7) demonstrates its advantage over all other evaluated architectures. This indicates that when the proposed Branchformer-based encoder and the proposed counting module work collaboratively, the model's overall performance in complex environments can be significantly improved.

\begin{table}
    \centering
    \begin{tabular}{ccccccc}
        \toprule
        \multirow{2}[0]{*}{Method} & \multirow{2}[0]{*}{FlexNumSpk} & \multicolumn{3}{c}{VoxCeleb2-2\&3mix} \\
          & & SI-SDRi & STOI  & SCA \\
        \midrule
        A4 & \CIRCLE & 8.64  & 0.57  & 59.87\% \\
        A5 & \CIRCLE & 9.08  & 0.59  & 64.62\% \\
        A6 & \CIRCLE & 10.45 & 0.68  & 75.79\% \\
        \textbf{A7} & \CIRCLE & \textbf{10.88} & \textbf{0.71} & \textbf{84.74\%} \\
        \bottomrule
    \end{tabular}
    \caption{Ablation study for A4--A7 structures. \CIRCLE denotes that the architecture is \textbf{able} to handle a flexible number of speakers.}
    \label{tab:ablation2}
\end{table}

\subsection{Quality Study}
\label{subsec:quality}
In this section, Quality Studies are conducted to comprehensively evaluate the robustness and generalization capabilities of the proposed model under diverse noisy scenarios and multi-speaker mixtures. The study also compares AV-DPRNN---a classical architecture for multi-speaker separation or extraction---and AV-Sepformer---the baseline backbone network of our model---under identical experimental settings.

\subsubsection{Quality Study for Noise Robustness}
Table \ref{tab:SNR} compares the performance of AV-DPRNN and the proposed AVFSNet on mixed speech with varying SNRs. Test sets were generated from the VoxCeleb2 dataset across SNRs ranging from -20 dB to 20 dB, with results stratified by SNR intervals. The proposed AVFSNet consistently outperforms AV-DPRNN across all SNR ranges, demonstrating robust noise tolerance in diverse environments. Notably, under extreme noise conditions (-20 dB to -10 dB), AVFSNet maintains viable separation quality. Moreover, the consistently high SI-SDR values achieved under high-SNR conditions indicate that speaker separation becomes significantly more tractable when the speech signal is less corrupted by noise.

\begin{table}
    \centering
    \begin{tabular}{cccccc}
        \toprule
        Method & Input SNR(dB) & SI-SDR & SDR   & SISDRi & SDRi \\
        \midrule
        \multirow{6}[0]{*}{AvDPRNN} & [-20, -10] & 0.11  & 1.28  & 15.14 & 14.62 \\
            & [-10, -5] & 6.66  & 7.43  & 14.16 & 14.57 \\
            & [-5, 0] & 9.27  & 9.94  & 11.77 & 12.28 \\
            & [0, 5] & 11.86 & 12.43 & 9.36  & 9.84 \\
            & [5, 10] & 14.47 & 14.96 & 6.97  & 7.39 \\
            & [10, 20] & 17.51 & 18.01 & 2.51  & 2.94 \\
        \midrule
        \multirow{6}[0]{*}{AVFSNet} & [-20, -10] & 4.73  & 5.94  & 19.76 & 19.28 \\
            & [-10, -5] & 9.41  & 10.35 & 16.91 & 17.48 \\
            & [-5, 0] & 11.82 & 12.66 & 14.32 & 15.01 \\
            & [0, 5] & 14.12 & 14.87 & 11.62 & 12.28 \\
            & [5, 10] & 16.31 & 17.01 & 8.81  & 9.43 \\
            & [10, 20] & 18.48 & 19.24 & 3.48  & 4.18 \\
        \bottomrule
    \end{tabular}
    \caption{Performance of AvDPRNN and the proposed AVFSNet on mixed speech datasets with different SNR}
    \label{tab:SNR}
\end{table}

\newpage

\subsubsection{Quality Study for Performance Extrapolation}
Table \ref{tab:QSNS} presents the performance of the proposed AVFSNet alongside AV-DPRNN and AV-Sepformer on mixed speech with varying numbers of speakers. AVFSNet achieves superior SI-SDRi and SDR across all tested speaker counts. Notably, even under extreme conditions with 10-speaker mixtures, the proposed method still achieves measurable SI-SDR improvement (2.69 dB).

\begin{table}
    \centering
    \begin{tabular}{llcccccc}
      \toprule
      \multicolumn{2}{c}{\multirow{2}{*}{Methods}} & \multicolumn{6}{c}{Number of Speakers}               \\
      \multicolumn{2}{c}{}                         & 2      & 3      & 4      & 6      & 8      & 10      \\
      \midrule
      \multirow{2}{*}{AvDPRNN}     & SISDRi        & 10.45 & 9.66  & 7.70  & 5.26  & 2.74  & 1.90   \\
                                   & SDR           & 11.26 & 7.09  & 3.32  & -1.13 & -4.93 & -6.71  \\
      \multirow{2}{*}{AvSepformer} & SISDRi        & 13.20 & 12.39 & 9.88  & 6.33  & 3.78  & 2.32   \\
                                   & SDR           & 13.84 & 9.62  & 5.25  & -0.31 & -4.24 & -6.24  \\
      \multirow{2}{*}{AVFSNet}     & SISDRi        & 14.00 & 13.39 & 10.85 & 6.89  & 4.09  & 2.69   \\
                                   & SDR           & 14.69 & 10.66 & 6.24  & 0.40  & -3.76 & -5.93  \\
      \bottomrule
    \end{tabular}
    \caption{Quality study results on model performance extrapolation.}
    \label{tab:QSNS}
\end{table}

\subsection{Visualization}
\label{subsec:visualization}
\begin{figure}
    \centering
    \includegraphics[width=0.9\linewidth]{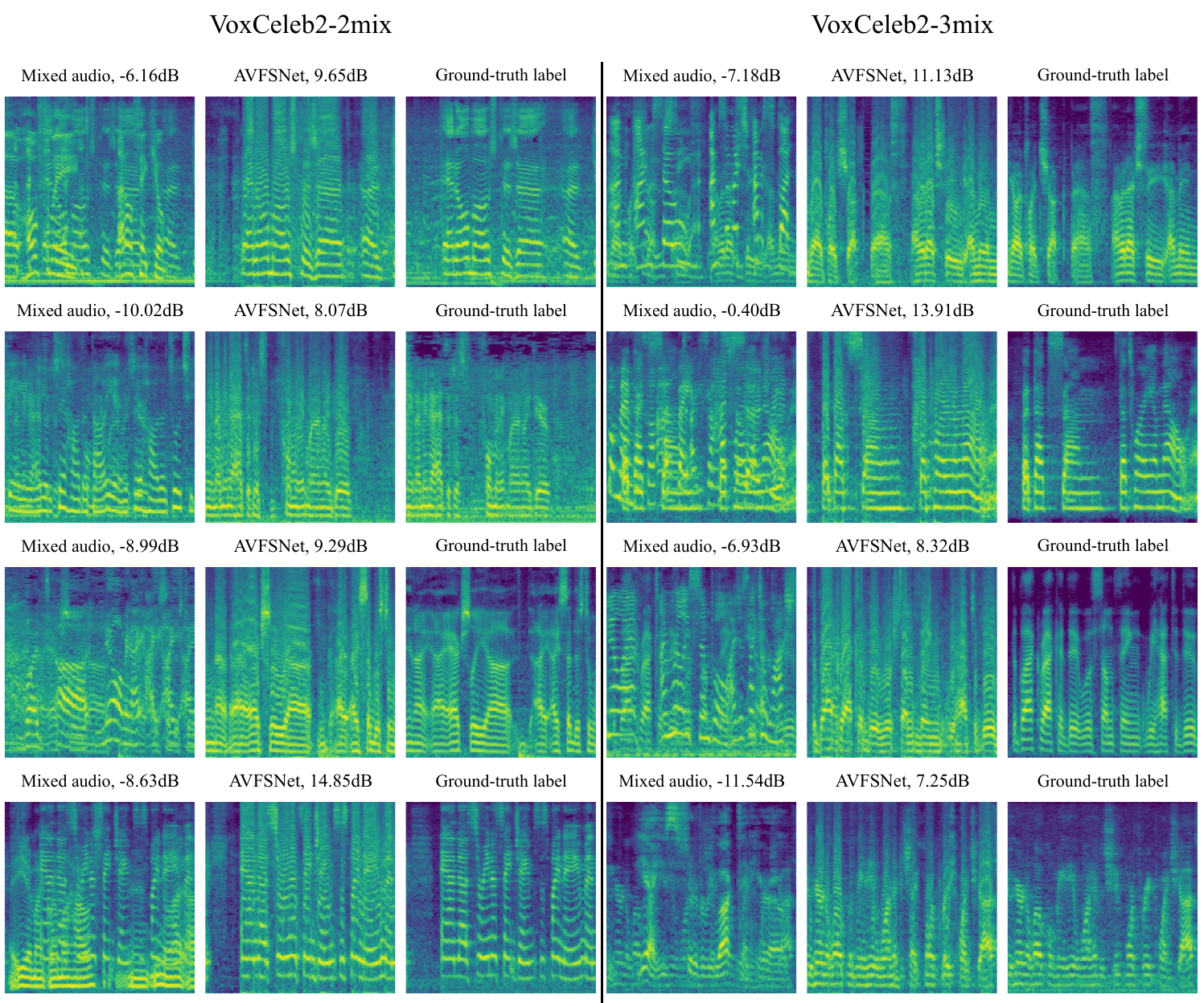}
    \caption{Visualization of 8 target speaker separation samples by AVFSNet on VoxCeleb2. Compared with mixed speech inputs, AVFSNet effectively filters environmental noise and achieves significant improvement in SI-SDR.}
    \label{fig:samplevisualization}
\end{figure}
In this section, we provide some visualization results of AVFSNet, implemented on the VoxCeleb2 dataset. Figure \ref{fig:samplevisualization} contains eight groups of audio sample extraction results, where each group includes:
\newpage
\begin{enumerate}
  \item Mel-spectrogram of the mixed speech
  \item Mel-spectrogram of AVFSNet-extracted speech
  \item Mel-spectrogram of Ground-truth speech
\end{enumerate}

In addition, the SI-SDR values are provided for both the mixed speech and AVFSNet-separated speech. The four left groups are derived from the two-speaker mixed speech dataset generated from VoxCeleb2, while the four right groups originate from the three-speaker mixed speech dataset. In each group, compared to mixed speech, AVFSNet outputs exhibit closer alignment with ground-truth speech and achieve significant SI-SDR improvements. These visualizations demonstrate that AVFSNet can achieve robust extraction capability across diverse scenarios (varying speaker counts and SI-SDR conditions).

\section{Conclusion}
In this paper, we propose AVFSNet, an audio-visual speech separation method for unknown-number-of-speakers scenarios. AVFSNet enhances model adaptability to complex acoustic environments by incorporating visual information and enabling joint learning of speaker counting and speech separation tasks. Furthermore, AVFSNet integrates multi-scale modeling with a parallel separation architecture, achieving superior separation performance. The experimental results demonstrate the superiority of AVFSNet, which outperforms previous methods across multiple datasets and diverse acoustic scenarios. In future work, we will further explore various extensions of the AVFSNet architecture and investigate more adaptive separation networks.

\section*{Acknowledgements}
This work was supported by the National Natural Science Foundation of China under Grant 62171263 and the Taishan Scholars Program.

\bibliographystyle{elsarticle-num}
\bibliography{./reference}
\end{document}